\documentclass[10pt, conference, letterpaper]{IEEEtran}
\IEEEoverridecommandlockouts

\usepackage{amsmath,amsthm,amssymb}
\usepackage{cite}
\usepackage{booktabs}
\usepackage{tikz}
\usetikzlibrary{calc}
\usepackage[binary-units]{siunitx}
\usepackage[shortlabels]{enumitem}
\setlist[enumerate,itemize]{nosep}
\usepackage{thmtools} \usepackage{placeins} \usepackage{float}
\usepackage{url}
\usepackage{balance}

\usepackage[noend,ruled,linesnumbered]{algorithm2e}
\DontPrintSemicolon
    
\SetKwFunction{Push}{Push}
\SetKwProg{fn}{Function}{:}{}
\SetKw{Break}{break}

\newcommand{\TabLabel}[1]{\label{tab:#1}}
\newcommand{\Table}[1]{Table~\ref{tab:#1}}

\newcommand{\FigLabel}[1]{\label{fig:#1}}
\newcommand{\Figure}[1]{Fig.~\ref{fig:#1}}

\newcommand{\SectLabel}[1]{\label{sect:#1}}
\newcommand{\Section}[1]{Sect.~\ref{sect:#1}}

\newcommand{\LemLabel}[1]{\label{lem:#1}}
\newcommand{\Lemma}[1]{Lemma~\ref{lem:#1}}
\newcommand{\PropLabel}[1]{\label{prop:#1}}

\newcommand{\ie}{i.e.}
\newcommand{\eg}{e.g.}
\newcommand{\etal}{et~al.}

\thickmuskip=\thinmuskip
\medmuskip=\thinmuskip
\newcommand{\N}{\ensuremath{\mathbb{N}}}

\newcommand{\Degree}{\ensuremath{\mathrm{deg}}}

\newcommand{\MaxDegree}{\ensuremath{\Delta}}

\newcommand{\NP}{NP}
\newcommand{\bigO}{\ensuremath{O}}

 \newcommand{\Demand}{\ensuremath{w}}

\newcommand{\B}{\ensuremath{k}}

\newcommand{\DjMatching}{\ensuremath{{\mathcal{M}}}}

\newcommand{\Coloring}{\ensuremath{\mathcal{C}}}
\newcommand{\Fan}{\ensuremath{\mathcal{F}}}
\newcommand{\fan}{f}

\newcommand{\AlgName}[1]{\texttt{#1}}
\newcommand{\SwapIn}[2]{\AlgName{SwapIn}(#1, #2)}
\newcommand{\SwapOut}[1]{\AlgName{SwapOut}(#1)}
\newcommand{\LocalSwaps}{\AlgName{LocalSwaps}}
\newcommand{\GreedyIt}[1]{\AlgName{GreedyIt#1}}
\newcommand{\misragries}{\AlgName{MG}}
\newcommand{\nodecentered}{\AlgName{NodeCentered}}
\newcommand{\NCthresh}{\ensuremath{\theta}}

\newcommand{\kEC}{\AlgName{kEC}}
\newcommand{\CCkEC}{\AlgName{CC}}
\newcommand{\RLkEC}{\AlgName{RL}}
\newcommand{\kECdoColor}{\AlgName{kColorEdge}}

\newcommand{\BatchGreedy}[1]{\AlgName{batch-greedy#1}}
\newcommand{\BatchNC}[1]{\AlgName{batch-NC#1}}
\newcommand{\BatchInvariant}{\AlgName{batch-apx}}
\newcommand{\dynkEC}[1]{\AlgName{dyn-\kEC{}#1}}
\newcommand{\hybridkEC}[1]{\AlgName{hybrid-\kEC#1}}
\newcommand{\hybridGreedy}[1]{\AlgName{hybrid-greedy#1}}
\newcommand{\dynGreedy}[1]{\AlgName{dyn-greedy#1}}
\newcommand{\GreedyRecDepth}{\ensuremath{\alpha}}
\newcommand{\GreedyRand}{\ensuremath{\beta}}

\newcommand{\MaxNumEdges}{M}
\newcommand{\MaxBatchSize}{B}
\newcommand{\Batch}{\ensuremath{\mathcal{B}}}
\newcommand{\BatchSize}{\ensuremath{b}}
\newcommand{\EUp}{\ensuremath{\mathcal{U}}}
\newcommand{\WeightChange}{\ensuremath{\delta}}
\newcommand{\NumUpdates}{\BatchSize}

\newcommand{\FilterThreshold}{t}

\newcommand{\DataSet}[1]{\texttt{#1}}
\newcommand{\DSfacebook}[1]{\if\relax\detokenize{#1}\relax \DataSet{FB}\else \DataSet{FB (#1)}\fi}
\newcommand{\DSfilterfacebook}[1]{\if\relax\detokenize{#1}\relax \DataSet{facebook-80}\else \DataSet{filtered-facebook (#1)}\fi}
\newcommand{\DSfb}[1]{\DataSet{faceb.}}
\newcommand{\DSffb}[1]{\DataSet{faceb.-f.}}
\newcommand{\DSfacebooksplit}{\DataSet{FB/s}}

\newcommand{\DSpfab}{\DataSet{pfab}}
\newcommand{\DShpc}{\DataSet{hpc}}

\newcommand{\AvgRunningTime}{\bar{\tau}}
\newcommand{\AvgSolutionWeight}{\bar{\sigma}}
\newcommand{\AvgColoringOps}{\bar{o}}
\newcommand{\Instance}{I}

\newcommand{\Alg}{\mathcal{A}}

\newcommand{\High}[1]{\textbf{#1}}
 \usepackage{graphicx}
 \usepackage{wrapfig}
\usepackage{calc}
\newcommand{\erclogowrapped}[1]{\setlength\intextsep{0pt}\begin{wrapfigure}[3]{r}{#1*\real{1.1}}\includegraphics[width=#1]{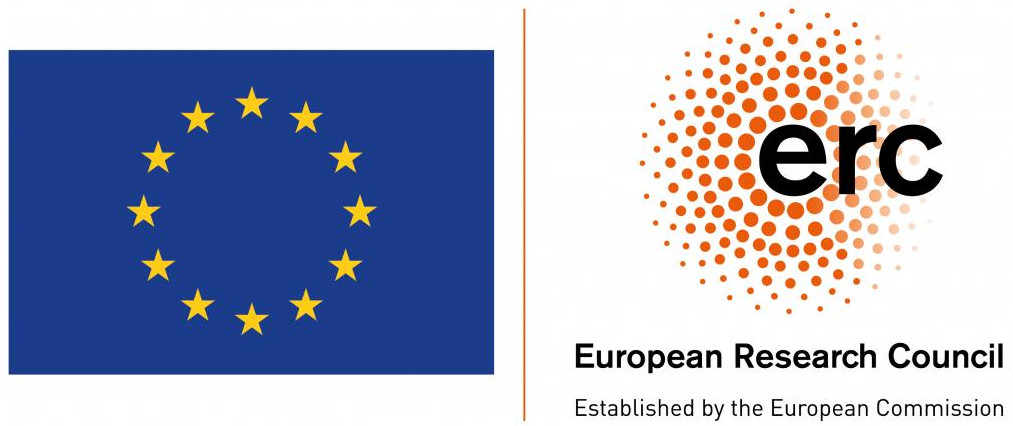}\end{wrapfigure}}

\usepackage{orcidlink} 

\usepackage{hyperref} \hypersetup{colorlinks, } 

\usepackage[normalem]{ulem}

\begin{document}

\title{Dynamic Demand-Aware Link Scheduling\\for Reconfigurable Datacenters
\thanks{\erclogowrapped{5\baselineskip}
This project has received funding from the European Research Council (ERC) under the European Union's Horizon 2020 research and innovation programme (MoDynStruct, No. 101019564) and the Austrian Science Fund (FWF) grant  \href{https://www.doi.org/10.55776/Z422}{DOI 10.55776/Z422}, grant  \href{https://www.doi.org/10.55776/I5982}{DOI 10.55776/I5982}, and grant  \href{https://www.doi.org/10.55776/P33775}{DOI 10.55776/P33775} with additional funding from the netidee SCIENCE Stiftung, 2020–2024.
The third author is funded by the Vienna Graduate School on Computational Optimization, FWF project no.\ W1260-N35.
}}
 
 \author{\thanks{\{kathrin.hanauer,monika.henzinger,lara.ost\}@univie.ac.at,
stefan.schmid@tu-berlin.de
\par © 2023 IEEE. Personal use of this material is permitted. Permission from IEEE must be obtained for all other uses, in any current or future media, including reprinting/republishing this material for advertising or promotional purposes, creating new collective works, for resale or redistribution to servers or lists, or reuse of any copyrighted component of this work in other works. }

 \IEEEauthorblockN{Kathrin Hanauer
\orcidlink{0000-0002-5945-837X} }
 \IEEEauthorblockA{\textit{Faculty of Computer Science}\\
 \textit{University of Vienna}\\
 \textit{Vienna, Austria}
 }
 \and
 \IEEEauthorblockN{Monika Henzinger
\orcidlink{0000-0002-5008-6530} }
 \IEEEauthorblockA{\textit{Faculty of Computer Science}\\
 \textit{University of Vienna}\\
 \textit{Vienna, Austria}
 }
 \and
 \IEEEauthorblockN{Lara Ost
\orcidlink{0000-0003-4311-9928} }
 \IEEEauthorblockA{\textit{Faculty of Computer Science}\\
 \textit{University of Vienna}\\
 \textit{Vienna, Austria}
 }
 \and
 \IEEEauthorblockN{Stefan Schmid
\orcidlink{0000-0002-7798-1711} }
 \IEEEauthorblockA{\textit{TU Berlin}\\
 \textit{Berlin, Germany }}
 }

\maketitle

\begin{abstract}
Emerging reconfigurable datacenters allow to dynamically adjust the network topology in a demand-aware manner. These datacenters rely on optical switches which can be reconfigured to provide direct connectivity between racks, in the form of edge-disjoint matchings. While state-of-the-art optical switches in principle support microsecond reconfigurations, the demand-aware topology optimization constitutes a bottleneck.

This paper proposes a dynamic algorithms approach to improve the  performance of reconfigurable datacenter networks, by supporting  faster reactions to changes in the traffic demand. This approach leverages the temporal locality of traffic patterns in order to update the interconnecting matchings incrementally, rather than recomputing them from scratch. In particular, we present six (batch-)dynamic algorithms and compare them to static ones. We conduct an extensive empirical evaluation on 176 synthetic and 39 real-world traces, and find that dynamic algorithms can both significantly improve the running time and reduce the number of changes to the configuration, especially in networks with high temporal locality, while retaining matching weight.
\end{abstract}


\section{Introduction}

The performance of many cloud-based applications critically depends on the underlying network, requiring high-throughput datacenter networks which provide extremely large bandwidth \cite{li2019hpcc,khani2021sip,mogul2012we}. For example,  in distributed machine learning and data mining applications that periodically require large data transfers, the network is increasingly becoming a bottleneck. High network throughput requirements are also introduced by today's trend of resource disaggregation in datacenters, 
where fast access to remote resources (e.g., GPUs or memory) 
is critical, and the trend to hardware-driven workloads such as distributed training \cite{li2019hpcc}.

The stringent throughput requirements of modern datacenter applications have led researchers to propose
innovative datacenter network designs which rely on \emph{reconfigurable topologies} \cite{zhou2012mirror,kandula2009flyways,hamedazimi2014firefly,osa,projector,hampson2021reconfigurable,douglis2021fleet,avin2020demand,teh2020flexspander,sigmetrics22cerberus}: emerging optical technologies allow to enhance existing datacenter networks with reconfigurable optical matchings. In particular, optical circuit switches can provide direct connectivity between datacenter racks, in the form of one \emph{matching} per optical circuit switch. This technology enables \emph{demand-aware networks} \cite{ccr18san}: the optical matchings, and hence the datacenter topology interconnecting the racks, are adjusted depending on the current communication traffic.

Such demand-aware topology optimizations are attractive as datacenter traffic typically features much spatial and temporal structure,
and most transmitted bytes belong to a small number of so-called \emph{elephant flows}~\cite{sigmetrics20complexity,Roy2015InsideTS,kandula2009nature}.
Thus, throughput may be significantly increased by optimizing the topology towards the current traffic matrix and elephant flows.
This can be achieved, for example, by providing direct connectivity between frequently communicating racks rather than serving traffic along multiple hops, which introduces a ``bandwidth tax''~\cite{mellette2017rotornet,sigmetrics22cerberus}.

\begin{figure}[tb]
\centering
\includegraphics[width=.85\linewidth]{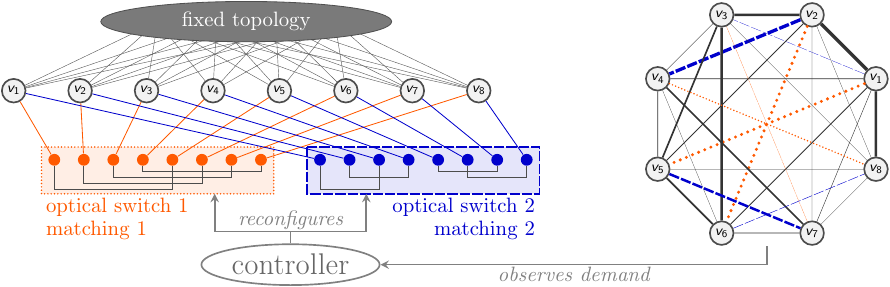}
\caption{Two optical switches establishing direct connections between nodes $v_1,\dots,v_8$, as instructed by the network controller, and the demand graph with the corresponding two matchings.}\FigLabel{datacenter}
\end{figure}

However, demand-aware topology optimizations are computationally expensive. 
In fact, state-of-the-art algorithms~\cite{projector,spaa21rdcn,helios,singla2010proteus,wang2010c,eclipse,apocs21renets} to compute optimized demand-aware switch matchings
for a given demand matrix can have a running time that is \emph{significantly higher} than the actual reconfiguration time provided by the state-of-the-art optical technologies (which is in the order of microseconds or even less~\cite{hanauer2022fast,ballani2020sirius}).
Thus, \emph{the computation of the demand-aware topology can be the bottleneck for demand-aware datacenter networks.}

Formally, the problem of optimizing the topology of a demand-aware network is a novel variant of 
a matching problem consisting of
``heavy'' disjoint matchings~\cite{hanauer2022fast} (see \Figure{datacenter}): given $k$ optical switches and a traffic demand (a.k.a.\ demand matrix) represented as a weighted graph where each node corresponds to a datacenter rack and  weighted edges represent demands, compute $k$ \emph{\mbox{(edge-)disjoint} matchings of high weight.}
The weight of these matchings hence corresponds to the amount of traffic which can be offloaded to the reconfigurable network, and hence to the throughput achieved by the datacenter network. 
In most existing reconfigurable datacenter architectures, including Helios~\cite{helios}, c-Through~\cite{wang2010c}, or recently Google's Gemini~\cite{zhang2021gemini}, 
among many others~\cite{spaa21rdcn,singla2010proteus,eclipse,apocs21renets}, 
this topology optimization is performed by a centralized software controller.
Computing such a set of matchings of maximum weight is \NP-hard and cannot be
approximated with arbitrary precision for all $k\geq2$~\cite{feige2002approximating,hanauer2022fast}.

The goal of this work is to improve the running time for computing $k$ disjoint matchings
to allow datacenter networks to react to changes in the demand more quickly and improve throughput further. 
Our main idea is as follows: Since traffic exhibits much temporal locality, it can be inefficient to \emph{recompute} the datacenter topology from scratch for each traffic matrix. Rather, we study \emph{dynamic algorithms}, i.e., algorithms which \emph{update} the topology incrementally, as a reaction to shifts in the demand matrix. Besides speeding up the computation, such dynamic algorithms may also require fewer changes in configurations to achieve a given throughput.
The latter benefits ongoing flows as fewer of them will be interrupted.

\begin{figure}[tb]
\centering
\includegraphics[width=.85\linewidth]{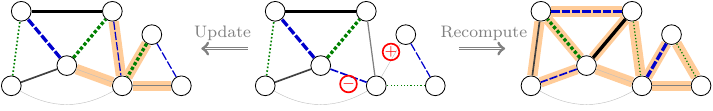}
\caption{Updating vs.~recomputing a solution from scratch for $\B=2$:
The two matchings are visualized in dotted green and dashed blue, the line width of an edge corresponds to its weight.
Shaded edges have been changed.
\\
\emph{Center:} Initial situation. Two updates arrive, one increases an edge weight
\protect\tikz[baseline=0pt]\protect\node[anchor=base,circle,red,draw,inner sep=.1pt,font=\small] {$+$};,
another decreases an edge weight
\protect\tikz[baseline=0pt]\protect\node[anchor=base,circle,red,draw,inner sep=.1pt,font=\small] {$-$};.
\emph{Left:} The result after processing the updates. Only four edges are affected.
\emph{Right:} The potential result after a full recomputation from scratch.
The whole network is reconsidered and also unaffected edges may have change, which can be inefficient.
}\FigLabel{motivation-dynamic}
\end{figure}
 We illustrate our motivation with a simple example network, see also
\Figure{motivation-dynamic}:
Assume the communication demand, i.e.~edge weight, of a node to two of its neighbors changes.
A \emph{dynamic} algorithm, which processes the updates itself, has the possibility to
adjust the configuration \emph{locally} and to leave most of the old solution
untouched.
By contrast, a \emph{static} one, which does a full recomputation from scratch, needs to
process the \emph{entire} network and, as it is unaware of the old configuration,
may additionally introduce unnecessary changes.
This can negatively affect both the
running time of the algorithm and the number of
changes to implement on the optical switches.

Thus, we study the following \emph{dynamic weighted $k$-disjoint matching problem}:
Given an undirected graph $G$ with edge weights representing the current
communication demands, process a sequence of \emph{batch updates}, each consisting of a set of (edge) updates,  where each (edge) update
either inserts or deletes an edge or changes the weight of one edge.
The main goal is to process each batch as quickly as possible and deliver an
up-to-date configuration, \ie, $k$ edge-disjoint matchings, after each batch
such that the total weight is maximized.
A secondary goal is to keep the \emph{recourse} small,
\ie, to minimize the number of changes to the matchings.

\paragraph*{Contributions}
We develop and evaluate a diverse set of algorithmic techniques for the weighted k-disjoint matching problem arising in the context of reconfigurable datacenter networks.
Based on the best static algorithms in~\cite{hanauer2022fast},
we design
two \emph{dynamic} algorithms,
which process each update individually,
three \emph{batch-dynamic} algorithms,
which process a batch of updates collectively,
and
two \emph{hybrid} algorithms,
which combine subroutines of static and dynamic approaches.
Furthermore, we introduce a \emph{universal speedup technique} that
filters out insignificant updates,
as well
as a \emph{universal post-processing routine} that
ensures an at least $\frac{1}{3}$-approximation of the maximum weight
and can also be run standalone.
We compare these algorithms and their speedup and postprocessing versions
in detail
with respect to 
solution weight, running time, and recourse
in theory and practice to the best static algorithms from~\cite{hanauer2022fast}.

\paragraph*{Main Experimental Results}
Our extensive study on \num{39} real-world and \num{176} synthetic instances shows that
our batch-dynamic and dynamic algorithms
can beat the best static algorithm w.r.t.~running time and recourse, while essentially retaining
the solution weight.
The combination of the post-processing and speedup technique proved to be particularly important
for the solution weight of the dynamic algorithms.
On instances where batches are small, their advantage over static algorithms is even
more pronounced.
Our two hybrid algorithms successfully combine the advantages of static and dynamic algorithms and are a good general-purpose choice.

We note that our approach is compatible with most existing reconfigurable datacenter architectures,
including~\cite{helios,wang2010c,spaa21rdcn,singla2010proteus,eclipse,apocs21renets,sigmetrics23duo,sigmetrics23mars},
which can hence directly benefit from these performance improvements.
Due to space restrictions, all proofs are omitted. 

\section{Preliminaries}\label{sec:prelim}
\paragraph{Basics}
We model the communication demand between peers as an undirected, weighted
graph $G = (V, E, \Demand)$ with node set $V$, edge set $E$, and
use $n := |V|$, $m := |E|$.
The function $\Demand: E \to \N_0$ maps each edge to a weight, which corresponds
to the respective communication demand.
An edge $e$ with $\Demand(e) = 0$ is considered \emph{absent}.
The maximum edge weight is $W := \max_{e \in E} \Demand(e)$.
The weight of a set of edges $S$ is 
$\Demand[S] = \sum_{e \in S} w(e)$.
Edges sharing an endpoint are \emph{adjacent}.
For two sets of vertices $X,Y \subseteq V$,
$E(X,Y) = E(Y,X) := \{\{x,y\} \in E \mid x \in X, y \in Y\}$ is the set of all
edges between $X$ and $Y$.
The
\emph{neighborhood} $N(e)$ of an edge $e$ is
$N(e) := E(\{u, v\}, V) \setminus \{e\}$.
Adding a set $S \subseteq E$ as subscript restricts
the neighborhood to  $N_S(e) := N(e) \cap S$.
The degree $\Degree(v)$ of a vertex $v$ is the number of edges incident to $v$,
and $\MaxDegree := \max_{v \in V} \Degree(v)$
is the maximum  degree.

A \emph{matching} in $G$ is a subset of $E$ such that no two edges are adjacent.
A \emph{$k$-disjoint matching} $\DjMatching$ is a set
of $k$ matchings $(M_1,\dots,M_k)$ such that $M_i \cap M_j = \emptyset$ for all
$i,j \in [k]$, $i \neq j$.
The \emph{weight} of a $k$-disjoint matching $\DjMatching$ is
$\Demand(\DjMatching) := \sum_{i=1}^k \Demand[M_i]$.
We can alternatively view a $k$-disjoint matching as a  \emph{partial (edge) $k$-coloring}.
Such a coloring $\Coloring: E \to [k] \cup \{\bot\}$ assigns to each edge one of $k$ colors or the symbol $\bot$,
where for each pair of adjacent edges $e, e'$ either $\Coloring(e) \neq \Coloring(e')$
or $\bot \in \{\Coloring(e), \Coloring(e')\}$.
An edge $e$ is \emph{uncolored} if $\Coloring(e) = \bot$ and \emph{colored} otherwise.
A color $c \in [k]$ is \emph{free} at vertex $v$ (resp.~edge $e$)
if there is no edge $e'$ incident to $v$ (resp.~adjacent to $e$) such that $\Coloring(e') = c$.
The set of all neighboring edges of an edge $e$ that have color $c$ is denoted by $N_c(e)$.
The \emph{weight} of a coloring $\Coloring$ is
$\Demand(\Coloring) := \sum_{e \in E: \Coloring(e) \neq \bot} \Demand(e)$,
i.e., the sum of the weights of all colored edges.
We mainly use the coloring perspective
and $k$ always refers to the number of colors, \ie~disjoint matchings.

\paragraph{Dynamic Setting}
Our algorithms maintain an partial $k$-coloring under \emph{edge updates},
where an (edge) update $\EUp(e, \WeightChange)$ is a change in the weight of an
edge $e$ by some amount $\WeightChange \neq 0$, \ie,
$\Demand(e) \gets \Demand(e) + \delta$.
The weight may either increase or decrease.
We consider a change from 0 to a positive weight  to be an \emph{insertion} and
a change from a positive weight to 0 to be a \emph{deletion}.
The updates are presented to the algorithm in a fixed order, and no
information about the updates themselves or the length of the update sequence
is known to the algorithms beforehand.
Additionally, the sequence of updates is partitioned into consecutive subsequences, each forming a \emph{batch}.
For each edge $e$ if there are multiple updates of $e$ in a batch, we add up the weight of the corresponding updates creating at most one update per edge per batch. 
The set of edges updated within a batch $\Batch$ is referred to as $E_\Batch$.
We assume that after each batch the algorithms are required to (i) output the value
of an up-to-date solution, and to (ii) be able to output the update to the solution
itself (\ie, the differences in the coloring) in time linear in the \emph{recourse}.
The \emph{recourse} corresponds to the total number of \emph{changes} in the coloring over a batch,
\ie, if $\Coloring'$ and $\Coloring$ are the edge colorings before and after a batch of updates $\Batch$, respectively, the recourse is $|\{e \in E_\Batch \mid \Coloring'(e) \neq \Coloring(e)\}|$.

We consider four types of algorithms:
\emph{Dynamic} algorithms  update their solution after every
individual update;
\emph{Batch-dynamic} algorithms process the entire batch at once and in self-chosen order;
\emph{Static} algorithms recompute from scratch after each batch.
\emph{Hybrid} algorithms decide on the basis of the observed updates whether they update the solution like a dynamic algorithm or recompute from scratch like a static one.

\paragraph{Related Work}
The study of the $k$-disjoint matching problem is motivated by emerging 
datacenter networks which augment a static (fixed) topology, 
with a dynamic topology implemented with $k$ optical (circuit) switches.
Each optical switch provides a set of exclusive direct connections between
top-of-rack switches that form a matching, For a recent survey of the field see Hall~\etal{}~\cite{osn21}.

Our work builds upon a recently suggested approach by
Hanauer~\etal{}~\cite{hanauer2022fast}, who showed that the static version of the problem is \NP-hard and no FPTAS can exist.
They evaluated a variety of algorithms in theory and practice.

The \emph{matching problem} has  been intensively studied in the \emph{dynamic} setting, see~\cite{HanauerH022} for a survey.
For the \emph{dynamic edge coloring} problem, 
only algorithms for the unweighted setting exist, which are not applicable to our setting.

Bienkowski \etal{}~\cite{DBLP:journals/sigmetrics/BienkowskiFMS20} studied a
problem in the online setting that is similar to $k$-disjoint matchings, but
can be solved to optimality in polynomial time.

\section{Algorithms}\SectLabel{algorithms}\label{sec:alg}

\begin{table}[tb]
  \caption{Running times and recourse for a batch of $\NumUpdates$ updates.\\
  $m$, $n$ and $\Delta$ refer to the cardinalities \emph{afterwards}. }\TabLabel{running-times}
\begin{tabular}{@{}lrr@{}}
    \toprule
    Algorithm
		& Batch Update Time
		& Recourse
		\\\midrule
    \GreedyIt{}      & $\bigO(m\log m + \B m)$                                                                    & $\bigO(m)$                      \\\nodecentered{}  & $\bigO(n\log n + m \log \MaxDegree + \B m)$                                                & $\bigO(m)$                      \\\kEC{}           & $\bigO(m\log m + kn^2)$                                                                    & $\bigO(m)$                      \\\midrule
    \dynGreedy{}\AlgName{($\GreedyRecDepth, \GreedyRand$)}$^*$ & $\bigO(\NumUpdates (\GreedyRand \cdot 2^\GreedyRecDepth + \GreedyRand^2))$                                           & $\bigO(\NumUpdates \cdot 2^\GreedyRecDepth)$  \\\dynkEC{}$^*$    & $\bigO(\NumUpdates\cdot n)$                                                                     & $\bigO(\NumUpdates\cdot n)$          \\\midrule
    \BatchGreedy{-l} & $\bigO(\MaxDegree \NumUpdates \log (\MaxDegree \NumUpdates) + k \MaxDegree^2 \NumUpdates)$ & $\bigO(\NumUpdates\cdot\MaxDegree)$ \\\BatchGreedy{}   & $\bigO(\MaxDegree \NumUpdates \log (\MaxDegree \NumUpdates) + k \MaxDegree \NumUpdates)$   & $\bigO(\NumUpdates\cdot\MaxDegree )$ \\\BatchNC{}       & $\bigO(\NumUpdates (\log \NumUpdates + \MaxDegree\log\MaxDegree + k\MaxDegree ))$          & $\bigO(\NumUpdates\cdot\MaxDegree )$ \\\BatchInvariant{}& $\bigO(m (k + \log m))$                                                                    & $\bigO(m)$                      \\\midrule
    \hybridGreedy{}\AlgName{($\GreedyRecDepth, \GreedyRand$)}
& \textit{see \Lemma{hybrid}} & $\bigO(m)$                      \\\hybridkEC{}     & $\bigO(m\log m + \B n^2)$                                                                    & $\bigO(m)$                      \\\midrule
\multicolumn{3}{l}{\footnotesize{$^*$for $\NumUpdates$ calls to \AlgName{AttemptMatch} or \AlgName{DecreaseWeight}}} \\
    \bottomrule
  \end{tabular}
\end{table}
 Our algorithms are in part based on static algorithms described in \cite{hanauer2022fast},
specifically \GreedyIt{}, \nodecentered{}, and \kEC{}.
For space reasons, we can only review these algorithms briefly. We then present our dynamic, batch-dynamic, and hybrid algorithms,
as well as a speedup technique and a postprocessing routine.
\Table{running-times} gives a compact overview over their asymptotic running time and recourse.

Below we use the two
functions
\High{\SwapIn{$e$}{$c$}} and
\High{\SwapOut{$e$}}, where $e$ is an edge and $c \in [\B]$ is a color.
\SwapIn{$e$}{$c$}
colors $e$ with $c$ and
uncolors $e$'s neighbors $N_c(e)$
if $\Demand(e) > \Demand[N_c(e)]$, otherwise it does nothing.
\SwapOut{$e$}
looks for up to two non-adjacent, uncolored edges $e'$ and $e''$ in $N(e)$ such that  (i)
$\Coloring(e)$ is free for both $e'$ and $e''$ and (ii) their total weight $w'$ is maximum over all such edge pairs.
If and only if $w' > \Demand(e)$, \SwapOut{$e$}
colors these edges with $\Coloring(e)$ and uncolors $e$.
\subsection{Static Algorithms}\SectLabel{static-algorithms}
We use the best static algorithms from  Hanauer
\etal{}~\cite{hanauer2022fast}, which we refer to for more details.
They are rerun from scratch after each batch of updates.
As they are ignorant of previous colorings, they
have a worst-case recourse of $\bigO(m)$.

The \High{\GreedyIt{}} algorithm is a ``greedy'' matching algorithm:
For each color $c$, all uncolored edges are processed in non-increasing order
of their weight and colored with $c$ if admissible.
At the end of the iteration for $c$, it optionally
performs a \High{\LocalSwaps{}} procedure, where
\SwapOut{$e$} is run on each edge of color $c$.

The \High{\nodecentered{}} algorithm
is also greedy, but proceeds node by node.
The nodes are processed in non-increasing order by their rating, which
corresponds to the sum of the weights of the $k$ heaviest incident edges.
For each node, up to $k$ incident edges are colored with any available color in
order of non-increasing weight.
To avoid an overly greedy coloring, the algorithm defers the coloring of edges
with weight below $\NCthresh \cdot W$, where $\NCthresh$ is a threshold parameter,
until after all nodes have been processed.
We use $\NCthresh = 0.2$, as suggested in~\cite{hanauer2022fast}.

The \High{\kEC{}} algorithm is based on the edge coloring algorithm of Misra and
Gries (\misragries)~\cite{DBLP:journals/ipl/MisraG92}.
Hanauer \etal{}~\cite{hanauer2022fast} adapted this algorithm to consider edge
weights and to limit the number of colors to $k$, and considered different
speedup techniques.
The key observations that allow a modification of the algorithm without
compromising its correctness are that \misragries{} (i) colors the edges
in arbitrary order and (ii) never uncolors an already colored edge.
We use \kEC{} as suggested~\cite{hanauer2022fast} with the flags \CCkEC{} and \RLkEC{} enabled.
{\kEC{}} tries to color the edges in non-increasing order
according to their weight, using a procedure \High{\kECdoColor},
which largely follows the routine in \misragries{}, as follows:
To color an edge $e = \{u,v\}$, {\kEC{}} first checks whether $u$ and $v$ both have
at least one free color each, and leaves $e$ uncolored otherwise.
If a \emph{common} free color exists, it is used for $e$ (\CCkEC{} flag).
Otherwise, following \misragries{}, \kEC{} constructs a fan $\Fan_u$ around $u$.
A \emph{fan} is a maximal sequence $\Fan_u = (\fan_0 = v, \fan_1, \dots, \fan_\ell)$ of distinct neighbors of $u$,
where for all $1 \leq i \leq \ell$, if $\{u,\fan_i\}$ has color $c_i$, then
color $c_i$ is free on $\fan_{i-1}$. 
Note that $\Fan_u$ is not necessarily unique.
Now we pick a color $c$ that is free at $u$ and a color $d$ that is free at $\fan_{\ell}$.
In \misragries{}, $d$ always exists, whereas in \kEC{}, the number of colors is  limited to $\B$.
Thus, if \kEC{} does not find a free color at $\fan_{\ell}$, \kECdoColor{} has failed for $u$ and \kEC{} repeats \kECdoColor{} symmetrically at $v$.
If \kECdoColor{} failed at both $u$ and $v$, $\{u,v\}$ remains uncolored.

Otherwise, assume that $\fan_{\ell}$ in fan $\Fan_u$ has a free color $d$.
We consider two cases.
(1) If $d$ is free at $u$,
\kEC{} \emph{rotates} the full fan $\Fan_u$ (\RLkEC{} flag),
\ie, for all $1 \leq i \leq \ell$, the edge $\{u,\fan_{i-1}\}$ is recolored with the color of $\{u,\fan_i\}$, 
and $\{u,\fan_{\ell}\}$ receives the color $d$.
(2)
Otherwise, as $d$ was free at $\fan_{\ell}$ and the fan is maximal,
there must be a neighbor $\fan_j$ in $\Fan_u$ such that $\{u, \fan_j\}$ has color $d$.
Like \misragries{}, \kEC{}
constructs
a path in the graph of maximal length that starts with the edge $\{u, \fan_j\}$ and whose colors alternate
between $d$ and $c$.
The colors $d$ and $c$ are then swapped along this path, which guarantees that $d$ is
free both at $u$
and at at least one vertex in $\Fan_u$~\cite{DBLP:journals/ipl/MisraG92}.
\kEC{} (and \misragries{}) now rotate $\Fan_u$ as in (1), but only up to the first
vertex $\fan_x$ where $d$ is free, and then recolor $\{u, \fan_x\}$ with $d$.

\subsection{Dynamic Algorithms}
The algorithm \High{\dynGreedy{}} is a dynamic enhancement of \GreedyIt{} and takes
two parameters $\GreedyRecDepth$ and $\GreedyRand$.
Each update is treated separately, where the subroutines \AlgName{AttemptColor}
and \AlgName{DecreaseWeight} handle weight \emph{increases} of \emph{uncolored} edges and weight \emph{decreases}
of \emph{colored} edges, respectively.
Nothing is done on weight increases of already colored edges or
weight decreases of uncolored edges.

\AlgName{AttemptColor}:
If there is a free color on $e$, we color $e$ with this color.
Otherwise, we try to find a color $c$ where
\SwapIn{$e$}{$c$} is successful.
If yes, we recursively call \AlgName{AttemptColor} on the
newly uncolored edges up to a recursion depth of $\GreedyRecDepth$.
To determine $c$, if $\GreedyRand \ge k$, $c$ is the
color of the edges adjacent to $e$ that have minimum total weight.
If $\GreedyRand< k$, we pick a subset of $\GreedyRand$
colors uniformly at random from $[k]$ and
just among these, we use the color where the edges adjacent to $e$ have the minimum total weight. 

\AlgName{DecreaseWeight}:
We first call a modified \SwapOut{$e$} which only considers a random subset of $\GreedyRand$ incident edges per end node.
If this changed the coloring, the algorithm tries to recolor $e$ using \AlgName{AttemptColor} with $\GreedyRecDepth = 0$
and $\GreedyRand$ as given.
The procedure is deterministic for $\GreedyRand = \MaxDegree$.

\begin{restatable}{lemma}{greedymatchingstime}\label{lem:greedy-matchings-time}
\dynGreedy{} processes an edge weight increase in time $\bigO(\GreedyRand \cdot 2^\GreedyRecDepth)$
and a decrease in $\bigO(\GreedyRand^2)$. The recourse is $\bigO(2^\GreedyRecDepth)$, respectively $\bigO(1)$. \end{restatable}

The dynamic $k$-edge coloring algorithm \High{\dynkEC{}} uses the \kECdoColor{} procedure of the static \kEC{} algorithm to color edges.
Similar to \dynGreedy{}, it tries to color a previously uncolored edge if its
weight increases, and the heaviest uncolored edges adjacent to an updated,
colored edge in case of a weight decrease.
Weight increases of colored edges and weight decreases of uncolored edges are again ignored.

Before calling \kECdoColor{} for an edge $\{u,v\}$,
the algorithm ensures that both $u$ and $v$ have at least one free color each (cf.~\Section{static-algorithms}):
Let $E_u = \emptyset$ if $u$ has a free color and let $E_u$ otherwise be the
singleton containing the colored edge incident to $u$ with the smallest weight.
$E_v$ is defined likewise for $v$.
If $E_u \cup E_v = \emptyset$, \kECdoColor{} is called immediately.
Otherwise, if $\Demand[E_u \cup E_v] < \Demand(\{u, v\})$,
the edges in $E_u \cup E_v$ are uncolored before \kECdoColor{} is called.
The uncoloring is undone if \kECdoColor{} failed to color $\{u,v\}$,
otherwise, the algorithm just tries to recolor an edge in $E_u \cup E_v$
with a free color.
No action is taken if $\Demand[E_u \cup E_v] \geq \Demand(\{u, v\})$.

\begin{restatable}{lemma}{dynkeclemma}
  \LemLabel{dyn-kec}
Algorithm \dynkEC{} processes each update in time $\bigO(n)$ and has a recourse of $\bigO(n)$.
\end{restatable}

\subsection{Batch-Dynamic Algorithms}\SectLabel{batch-greedy}
The algorithm \High{\BatchGreedy{}} is a batch-dynamic adaptation of the static \GreedyIt{} algorithm. For each batch, it processes only those edges that are present and either updated or adjacent to at least one updated or deleted edge.
The latter ensures that an edge with large weight increase can be colored at the expense of an unchanged neighboring edge.
The edges to be processed are first uncolored
and then treated like the set of all edges in \GreedyIt{},
including the optional \LocalSwaps{} procedure.
We refer to the algorithm by \BatchGreedy{-l} with \AlgName{LocalSwaps}
and by \BatchGreedy{} without.

\begin{restatable}{lemma}{batchiterativegreedytime}\label{lem:batch-iterative-greedy-time}
For a batch of size $\BatchSize$,
	\BatchGreedy{}
	and
	\BatchGreedy{-l}
	run in time
	$\bigO(\NumUpdates\MaxDegree \log (\NumUpdates\MaxDegree) + k \NumUpdates \MaxDegree)$
	and
	$\bigO(\NumUpdates \MaxDegree \log (\NumUpdates \MaxDegree) + k \NumUpdates \MaxDegree^2)$,
	respectively.
The recourse is $\bigO(\NumUpdates \MaxDegree)$.
\end{restatable}

The batch node centered algorithm \High{\BatchNC{}} is
an adaptation of the static \nodecentered{} algorithm and,
similar to \BatchGreedy{},
processes only updated edges and their adjacent edges.
For every node that is incident to an updated edge, its rating is computed as in
\nodecentered{} and \emph{all} its incident edges are uncolored.
These nodes are then ranked and processed as in \nodecentered{}, using the same
threshold parameter $\NCthresh = 0.2$.

\begin{restatable}{lemma}{batchnodecenteredtime}\label{lem:batch-node-centered-time}
For a batch of size $\BatchSize$,
\BatchNC{} runs in time $\bigO(\NumUpdates (\log \NumUpdates + \MaxDegree\log \MaxDegree + k\MaxDegree))$.
The recourse is $\bigO(\NumUpdates\MaxDegree)$.
\end{restatable}

\subsection{Hybrid Algorithms}\SectLabel{hybrid}
If a large fraction of the edges in the graph is updated, it can be better to
recompute a coloring from scratch instead of processing each update in the
batch individually.
To this end, we combine the static algorithm with the best time-for-weight
tradeoff~\cite{hanauer2022fast}, \kEC{}, with the update procedures of
our two dynamic algorithms, \dynGreedy{} and \dynkEC{},
and thus obtain a \High{\hybridGreedy{}} and a \High{\hybridkEC{}} algorithm.

The hybrid algorithms choose between the static and dynamic algorithm with the
goal to minimize the running time.
For the decision, they use a simple heuristic that can be computed quickly
and relates the size of the batch $\BatchSize$ to $n$:
if $\BatchSize < n$, the dynamic algorithm should be faster,
otherwise, we expect the dynamic algorithm to do much more work than
the static, also due to bookkeeping, so the static should be faster.
To run the procedures of the dynamic algorithms at the time the update is
observed, the decision needs to be made \emph{before} the next batch
arrives.
We hence use the batch size $\BatchSize'$ of the \emph{previous} batch as an estimate for the current batch.

\begin{restatable}{lemma}{hybridlemma}\LemLabel{hybrid}
\hybridGreedy{},
parameterized by $\GreedyRecDepth$ and $\GreedyRand$,
and
\hybridkEC{}
process a batch of updates in
$\bigO(n \cdot (\GreedyRand \cdot 2^\GreedyRecDepth + \GreedyRand^2) + m\log m + \B n^2)$
and $\bigO(m\log m + \B n^2)$ time, respectively.
The recourse is $\bigO(n)$.
\end{restatable}

\subsection{Reducing Work by Filtering Updates}\SectLabel{update-filtering}

Minor changes in the weight of an edge do not have a large effect on the overall solution weight.
This holds particularly in large graphs, where the contribution of an individual edge to the solution weight is small by comparison.
We filter such updates with the goal of improving the running time without degrading the weight of the solutions.

Our strategy is parameterized by a threshold $\FilterThreshold{} \geq 1$. Let $\EUp(e, \WeightChange)$ be an update that changes the weight of $e$ from $w$ to $w' = w + \WeightChange$.
With the filtering enabled, updates with $w, w' \neq 0$ and $\frac{w'}{w} \in [1/\FilterThreshold{},\FilterThreshold{}]$ are not processed by the dynamic algorithms. Insertions and deletions are never discarded.
We suffix an algorithm with \High{\AlgName{-f}} if filtering is used.

\subsection{Post-Processing and Approximation Guarantees}\SectLabel{postprocessing-approx}
To improve the weight of a given coloring, we consider a
post-processing routine that performs local optimization in the neighborhood of uncolored edges.
It ensures that for every color $i$, an uncolored edge has either one or two edges
with color $i$ in its neighborhood that are (in sum) at least as heavy as the uncolored
edge.
It establishes the following invariant:
\begin{restatable}{invariant}{ppinvariant}\PropLabel{postprocessing-invariant}
For every uncolored edge $e \in E$ and each color $c \in [\B]$,
		$\Demand[N_c(e)] \geq \Demand(e)$.
\end{restatable}

The post-processing algorithm places all uncolored edges in a priority queue, ordered by decreasing edge weight.
For each edge $e$ that is removed from the queue, it
(1) colors it with a free color $c$ if one exists, or (2) finds a color $c'$ for which the invariant is violated and performs a \SwapIn{$e$}{$c'$}, which must succeed due to the violation.
The newly uncolored edges are then pushed onto the queue for an invariant check.
Due to the violated invariant, they must both be strictly lighter than $e$.
Thus, each edge can be enqueued and dequeued at most once
and the procedure terminates after at most $m$ iterations.
(3) If the invariant is satisfied for all colors, $e$ remains uncolored.

The post-processing can be combined with any of the aforementioned algorithms by applying it to the coloring after a batch is complete.
We suffix an algorithm with \High{\AlgName{-p}} if post-processing is used.

\begin{restatable}{lemma}{twoapproxlemma}\LemLabel{twoapprox}
The weight of a $k$-disjoint matching $\DjMatching$ satisfying Invariant~\ref{prop:postprocessing-invariant}
    is at least $\frac{1}{3}$ of the weight of an optimal $k$-disjoint matching for $k > 1$.
    For $k = 1$ this factor is $\frac{1}{2}$.
\end{restatable}
\begin{proof}
    Let $\DjMatching^* =
    (M^*_1,M^*_2,\dots,M^*_k)$ be an optimal $k$-disjoint matching,
    and let $\DjMatching = (M_1,M_2,\dots,\allowbreak M_k)$ be a $k$-disjoint matching that satisfies Invariant~\ref{prop:postprocessing-invariant}.
    Let $\Coloring^*$ and $\Coloring$ be the colorings corresponding to $\DjMatching^*$ and $\DjMatching$, respectively.

    The weights of $\DjMatching^*$ and $\DjMatching$ can be expressed in terms of the weights of edges that
    \begin{enumerate}
        \item have \emph{equal} color in both solutions,
        \item are \emph{colored} in $\DjMatching$, but not in $\DjMatching^*$,
        \item are \emph{uncolored} in $\DjMatching$, but colored in $\DjMatching^*$,
        \item are colored in both solutions, but with \emph{different} colors.
    \end{enumerate}
    We define the corresponding sets
    \begin{enumerate}
        \item $E_= := \{e \in E \mid \Coloring^*(e) = \Coloring(e) \neq \bot \}$,
        \item $E_c := \{e \in E \mid \Coloring^*(e) = \bot \wedge \Coloring(e) \neq \bot \} = \DjMatching\setminus\DjMatching^*$ and \\
              $E_u := \{e \in E \mid \Coloring^*(e) \neq \bot \wedge \Coloring(e) = \bot \} = \DjMatching^*\setminus\DjMatching$,
        \item $E_{\neq} := \{e \in E \mid \Coloring^*(e) \neq \Coloring(e) \wedge \Coloring^*(e) \neq \bot \wedge \Coloring(e) \neq \bot \}$.
    \end{enumerate}
    Then, $\Demand(\DjMatching^*) = \Demand(E_=) + \Demand(E_u) + \Demand(E_{\neq})$
    and $\Demand(\DjMatching) = \Demand(E_=) + \Demand(E_c) + \Demand(E_{\neq})$.

    Consider an edge $e \in E_u = \DjMatching^* \setminus \DjMatching$.
    Let $i \in [\B]$ such that $e \in M^*_i$.
    As $e \not\in\DjMatching$, there is either one adjacent edge $e' \in N(e) \cap M_i$
    with $\Demand(e') \geq \Demand(e)$
    or two adjacent edges $e', e'' \in N(e) \cap M_i$
    with $\Demand(e') + \Demand(e'') \geq \Demand(e)$
    by Invariant~\ref{prop:postprocessing-invariant}.
    We charge $e'$ a cost of $\Demand(e)$ in the first case
    and $e'$ and $e''$ together a cost of $\Demand(e)$ in the second case,
    such that the respective cost shares do not exceed $\Demand(e')$ and $\Demand(e'')$.
    Since $e',e'' \in M_i$ and $e',e'' \notin M^*_i$ it holds that $e',e'' \in E_c \cup E_{\neq}$.

    Now consider an edge $e \in E_c = \DjMatching\setminus\DjMatching^*$ and let $j \in [\B]$ such that $e \in M_j$.
    Then, there can be at most two edges $e', e'' \in N(e)$ that are contained in $M^*_j$, but not in $\DjMatching$,
    so $e$ can be charged at most a cost of $2\Demand(e)$.
    
    Similarly, consider an edge $e \in E_{\neq}$ with $e \in M_j$.
    Again, there can be at most two edges $e', e'' \in N(e)$ that are contained in $M^*_j$, but not in $\DjMatching$,
    and $e$ is again charged at most a cost of $2\Demand(e)$.

    Thus, each edge in $e \in E_u$ with $\Coloring^*(e) = i$ is charged to an edge in $(E_c \cup E_{\neq}) \cap M_i$,
    and to each edge $e' \in E_c \cup E_{\neq}$ we charge a cost of at most $2\Demand(e')$.
    This implies $\Demand(E_u) \leq 2\Demand(E_c) + 2\Demand(E_{\neq})$
    and it follows that 
    \begin{equation*}
        \begin{aligned}
            \Demand(\DjMatching^*) &= \Demand(E_=) + \Demand(E_u) + \Demand(E_{\neq}) \\
                                   &\leq \Demand(E_=) + 2\Demand(E_c) + 2\Demand(E_{\neq}) + \Demand(E_{\neq}) \\
                                   &\leq 3\Demand(\DjMatching).
        \end{aligned}
    \end{equation*}
    If $k=1$ then $E_{\neq} = \emptyset$ and $\Demand(\DjMatching^*) \leq 2\Demand(\DjMatching)$.
\end{proof}
\begin{restatable}{corollary}{postprocesscorollary}
    The post-processing algorithm outputs a $\frac{1}{3}$-approximation for the weighted $k$-disjoint matching problem for $k > 1$.
    For $k=1$ it outputs a $\frac{1}{2}$-approximation.
\end{restatable}
\begin{proof}
    The post-processing algorithm outputs a $k$-disjoint matching that satsifies Invariant~\ref{prop:postprocessing-invariant}.
    The claim follows by \Lemma{twoapprox}.
\end{proof}

The post-processing can also be run on its own as a batch-dynamic algorithm, which we refer to as \High{\BatchInvariant{}}.
This algorithm collects those edges for which the invariant may have been violated by an update.
At the end of a batch, it processes the collected edges as described above.

\begin{restatable}{lemma}{pptime}
The post-processing algorithm and \BatchInvariant{} run in
		time $\bigO(m(\B + t_{+} + t_{-}))$,
    where $t_{+}$ and $t_{-}$ are the times for enqueuing and dequeuing an edge.
		The recourse is $\bigO(m)$.
\end{restatable}

\section{Experiments}\label{sec:eval}
We evaluate the running time, solution weight, and recourse of the algorithms from
\Section{algorithms} in practice on a large and diverse set of instances for
$k \in \{2,4,8,16,32\}$.
\subsection{Instances}\SectLabel{instances}
To perform a thorough analysis we include both
\emph{real-world} as well as large \emph{synthetic} instances.
All real-world instances were generated from measured or simulated network traffic on real-world networks.
Synthetic instances are random dynamic weighted graphs generated according to different models. 

\paragraph{Real-World Instances}
\begin{table}[tb]
    \caption{Real-world and split instances:
    max.\ \#edges~$\MaxNumEdges$ at any point in time, max.\ batch size~$\MaxBatchSize$,
    max.\ \#batches~$b$,
    max.\ \#nodes~$n$, percentage of insertions $I$, deletions $D$ and
    weight changes $C$ among updates, and the number of instances in each dataset.}\TabLabel{instance-summary}
    \centering
\addtolength{\tabcolsep}{-2pt}
\renewcommand{\arraystretch}{.8}
    \begin{tabular}{lrrrrrrrr}
      \toprule
      dataset    & $M$ & $B$ & $b$ &$n$ & $I$ & $D$ & $C$ & \#\\
      \midrule
      \DSfacebook{}       & \num{66423} & \num{66421} & \num{1399} & \num{367 } & \SI{6}{\percent} & \SI{4}{\percent} & \SI{90}{\percent} & \num{9 } \\
      \DShpc{}            & \num{9330}  & \num{16755} & \num{5000} &\num{1024} & \SI{36}{\percent} & \SI{36}{\percent} & \SI{28}{\percent} & \num{12} \\
      \DSpfab{}           & \num{1390}  & \num{2588}  & \num{4999} &\num{144 } & \SI{33}{\percent} & \SI{33}{\percent} & \SI{33}{\percent} & \num{9 } \\
      \DSfacebooksplit{} & \num{62799} & \num{64777} & \num{27980} & \num{367} & \SI{48}{\percent} & \SI{47}{\percent} & \SI{5}{\percent} & \num{26} \\
      \bottomrule
    \end{tabular}
\end{table}

We obtained dynamic instances from three real-world data collections that have been
used to analyze network communication before~\cite{DBLP:journals/sigmetrics/BienkowskiFMS20,hanauer2022fast}.

The \DataSet{facebook} (\DSfacebook{})~\cite{Roy2015InsideTS} datasets consist of IP packet information from Database, WebService and Hadoop clusters.
Each packet is associated with a Unix timestamp, the packet size, and the source and destination server.
We construct our instances as follows:
(1)~We group $x$ timestamps into one batch of updates.
(2)~For each source-destination pair $(a,b)$, we sum the size of all packets occurring in this batch,
which defines the new weight of the edge $(a,b)$ in the graph.
This may result in an edge insertion or deletion if the old or new weight, respectively, is zero.
From each of the three clusters we obtain three instances by choosing $x \in \{\num{60}, \num{1800}, \num{3600}\}$.

The three \DSpfab{}~\cite{sigmetrics20complexity,Alizadeh2013pFabricMN} and four \DShpc{}~\cite{sigmetrics20complexity} datasets consist of source-destination pairs,
each of which
is associated with a unique sequence number.
We proceed similarly to the \DSfacebook{} dataset, with the sequence number as timestamp and the number of packets per batch as the packet size and edge weight.
We obtain three instances per dataset by choosing the group size $x \in \{\num{10}, \num{100}, \num{1000}\}$.

\paragraph{Split Instances}The \DSfacebook{} instances are based on traces with a temporal resolution of one second.
To simulate instances with higher resolution and more frequent reconfigurations of the optical network as well as to study the influence of edge weights,
we generate a set of \emph{split instances} as follows: We distribute the weight of an edge in a batch over multiple ``sub-batches''
such that its new weight never exceeds a threshold
$z \in \{
\num{e5},
\num{e6}
\}$ and set the number of sub-batches that are created from each original
batch to $y \in \{5,10,15,20\}$.
For an original batch $\Batch$, we
create exactly $y$ sub-batches
$\Batch_1, \dots, \Batch_y$ and, for each edge that is updated to weight $w$ in
$\Batch$, we randomly select $\lceil\frac{w}{z} \rceil$ consecutive sub-batches
from $\Batch_1, \dots, \Batch_y$
and create the corresponding update operations.
For example, for $y=5$, and $z=\num{e5}$,
if there is an edge $e$ that is updated to $w=\num{215342}$ in $\Batch$, 
$e$ will have weight $z$ during two sub-batches and the remaining weight $r =\num{15342}$
in a third.
If the three randomly chosen consecutive sub-batches are $\Batch_2, \Batch_3, \Batch_4$,
we create update operations that set $e$'s weight to $z$ in $\Batch_2$,
reduce it to $r$ in $\Batch_4$, and to weight $0$ in $\Batch_5$ (\emph{zeroing}),
which means to delete $e$.
Note that $e$ keeps its weight of $z$ during $\Batch_3$, so no update is required.
The zeroing is omitted if the last sub-batch is among the randomly chosen ones
and $e$'s weight is updated in the first sub-batch $\Batch'_1$
of the following (original) batch $\Batch'$.
Thus, we create at most three updates for each update in the original instance.
As a split instance has $y$ times as many batches as the
one it was created from, the batch sizes
in the split instances are decreased.
Smaller values for $z$ increase the number of sub-batches $\lceil\frac{w}{z} \rceil$
that need to be selected, which reduces the number of zeroings
and further decreases the batch sizes.
For each combination of $s$ and $l$ and each instance in \DSfacebook{},
we create a split instance, but only if its maximum edge weight is at most $y \cdot z$.
The resulting data set \DSfacebooksplit{} contains \num{26} instances.

\Table{instance-summary} summarizes the properties of the real-world and split instances. On average over the batches,
\SI{47}{\percent} of the updates in \DSfacebook{}, \DShpc{}, and \DSpfab{} are weight changes, \SI{28}{\percent} are insertions, \SI{25}{\percent} are deletions and \SI{130}{\percent}\footnote{W.r.t.\ the number of edges \emph{after} the batch. Deletions can cause values above \SI{100}{\percent}.}
of the edges are updated.
Splitting the batches and distributing the updates over sub-batches in \DSfacebooksplit{} greatly increases
the relative number of insertions and deletions.
\Figure{weights} shows the weight distributions.

\begin{figure}
    \centering
    \includegraphics[width=.85\linewidth]{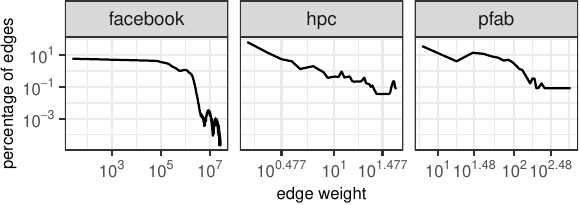}\caption{Weight distributions of the four datasets on a log-log scale.}\FigLabel{weights}\FigLabel{complexity}
\end{figure}

\paragraph{Synthetic Instances}
To diversify our test set
and to further study the
algorithms' behavior on large instances, we generated synthetic dynamic graphs
from RMAT~\cite{DBLP:conf/sdm/ChakrabartiZF04} instances, which have also been used
to evaluate the static algorithms~\cite{hanauer2022fast}.
Following~\cite{hanauer2022fast},
the initiator matrices for the RMAT generator are
$(0.55, 0.15, 0.15, 0.15)$ (\DataSet{rmat\_b}),
$(0.45, 0.15, 0.15, 0.25)$ (\DataSet{rmat\_g}), and
$(0.25, 0.25, 0.25, 0.25)$ (\DataSet{rmat\_er}),
with a number of nodes $n = 2^x$ and $14 \leq x \leq 18$.
Whereas each edge is equally likely for the \DataSet{rmat\_er} (Erdős-Rényi) graphs, \DataSet{rmat\_b} and
\DataSet{rmat\_g} instances have skewed normal degree distributions,
small-world properties, and larger clustering coefficients~\cite{DBLP:journals/ijhpca/HalappanavarFVTP12}.
The weights follow an exponential distribution with values between \num{1} and \num{500000}.
We create the dynamic instances as follows:
In the first batch, all edges of the static graph are inserted.
In every subsequent batch, a fraction $f \in \{0.1, 0.3, 0.5, 0.7, 0.9\}$ of edges is updated.
If the edge currently has positive weight, its weight is set to zero with probability $p \in \{0.1, 0.3\}$, i.e., the edge is deleted.
Otherwise we assign a new weight uniformly at random from the weights in the original instance.
We obtain \num{150} instances altogether,
which we group by graph size $n$.
Every group contains \num{30} instances ($3 \times 5 \times 2$),
each of which has \num{30} batches of updates.

On average over all synthetic instances, a batch consists of \SI{17}{\percent} insertions, \SI{16}{\percent} deletions and \SI{67}{\percent} weight changes.

\subsection{Setup and Methodology}\SectLabel{evaluation}
We
implemented\footnote{Source code available at~\url{https://github.com/DJ-Match/DyDJ-Match}~\cite{code}.} the algorithms in C++17 on top of the Algora library\footnote{\url{https://libalgora.gitlab.io}}.
We record the running time and weight of the solution for each batch of updates.
When applying a batch, the Algora framework notifies the algorithms of every update individually.
The dynamic algorithms \dynGreedy{} and \dynkEC{} immediately update the disjoint matching;
the batch-dynamic algorithms \BatchGreedy{}, \BatchNC{}, and \BatchInvariant{} use these notifications to prepare the batch,
i.e., to build up the sets of (nodes incident to) updated edges.

We obtain the number of changes to the coloring over a batch by comparing the coloring before and after the batch.
This is done by explicitly storing the state of the coloring before the batch and counting the changes in a separate run,
where no running times are measured.

We run our experiments on a machine (\texttt{A}) with two Intel Xeon Gold 6130 CPUs (16 cores each) and \SI{256}{\giga\byte} of main memory,
and a machine (\texttt{B}) with two Intel Xeon E5-2643 CPUs and
$2\times$\SI{750}{\giga\byte} of main memory.
Machine \texttt{A} was used for all experiments on real-world and synthetic
instances, whereas experiments for the split instances were run on machine \texttt{B}.
For each experiment, the process was pinned to a NUMA node and
its local memory.
We focus on the time to obtain a new configuration,
\ie, the running time of our algorithms, which are designed to be run on a central server.

To obtain reliable running times, we repeated each experiment three times.
For the analysis,
we take the median time over each batch for the deterministic algorithms
and
the arithmetic mean for the randomized algorithms, using different seeds.
For the \emph{average per-update time} $\AvgRunningTime_{\Instance}(\Alg)$ of an algorithm $\Alg$ on an instance $\Instance$,
we divide the time per batch by the batch size and take the arithmetic mean over all batches.
The \emph{average solution weight} $\AvgSolutionWeight_{\Instance}(\Alg)$
is the arithmetic mean over all batches.
In case of deterministic algorithms, the \emph{average recourse} $\AvgColoringOps_{\Instance}(\Alg)$
is defined analogously, whereas for randomized algorithms,
we first take the mean recourse per batch over all repetitions and then
the mean over these means.

  We compare the algorithms relative to each other w.r.t.\ speedup, solution weight, and recourse.
When comparing algorithm $\Alg$ to a reference algorithm $\mathcal{R}$ for an instance $\Instance$, we obtain
	(1) the \emph{speedup} as $\frac{\AvgRunningTime_{\Instance}(\mathcal{R})}{\AvgRunningTime_{\Instance}(\Alg)}$,
	(2) the \emph{relative solution weight} as $\frac{\AvgSolutionWeight_{\Instance}(\Alg)}{\AvgSolutionWeight_{\Instance}(\mathcal{R})}$,
	(3) the \emph{relative recourse} as $\frac{\AvgColoringOps_{\Instance}(\Alg)}{\AvgColoringOps_{\Instance}(\mathcal{R})}$.
  To average over a dataset, we always use the geometric mean.
  Observe that $\Alg$ is better than $\mathcal{R}$ in terms of speedup and relative solution weight if the value is greater than \num{1.0} and vice-versa for recourse.

\subsection{Results}
We analyze our algorithms first in smaller groups on the set of real-world instances,
then consider their performance in terms of running time, solution weight, and recourse
on the split and synthetic instances.

\paragraph*{Filtering Updates}
The update filtering strategy described in \Section{update-filtering} aims at ignoring small updates that are not expected to lead to large changes in the coloring.
A preliminary parameter study demonstrated that 
the decrease in running time due to filtering roughly matched the decrease in
solution weight.
Combined with post-processing, the speedup was retained, but the solution weight was not impaired.
Hence, we present only results for filtering together with post-processing.

\paragraph*{Parameters for \dynGreedy{}}
Preliminary results showed that increasing the recursion-depth parameter $\GreedyRecDepth$
beyond $\GreedyRecDepth = 1$
yields no significant improvement in the solution weight, but increases the running time.
Similarly, when comparing the randomized version of \dynGreedy{} with $\GreedyRand < \max\{\B,\MaxDegree\}$ to the deterministic version
with $\GreedyRand = \max\{\B,\MaxDegree\}$,
we observed that setting $\GreedyRand = 1$ yields the best trade-off between running time and solution weight for the randomized algorithm. 
In the following we thus consider only two versions:
a randomized one with $\GreedyRand = 1$, which is identified by the suffix \High{\AlgName{-r}},
and a deterministic one without this suffix.

\begin{figure}
    \centering
    \includegraphics[width=.75\linewidth]{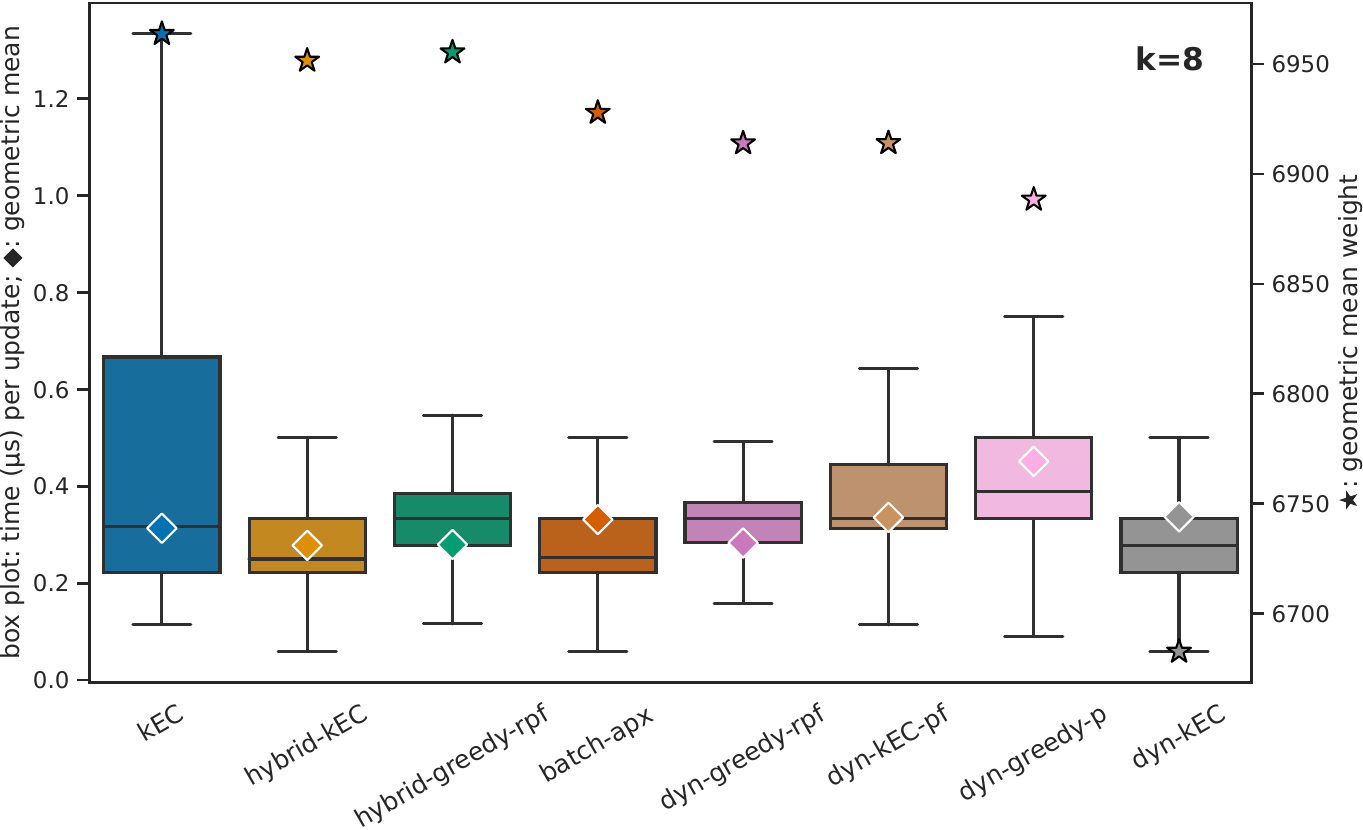}
    \caption{Average per-update times (left axis, boxes and $\blacklozenge$) and weight (right axis, $\star$) for $\B=8$ on all real-world instances.}\FigLabel{rw-boxplot-tpu}
\end{figure}
\subsubsection{\underline{Real-World Instances}}
\Figure{rw-boxplot-tpu} provides an overview over per-update times and solution weights for selected algorithms.

\paragraph*{Batch-Dynamic and Static Algorithms}
We evaluate the performance of these algorithms relative to each other.
Our results \emph{confirm} the earlier study for the static setting~\cite{hanauer2022fast}
and show that \kEC{} is on average the \emph{fastest static} algorithm.
The same applies to
\BatchGreedy{}, \BatchGreedy{-l}, and \BatchNC{}.
For all instances and all $\B$, the speedup of \nodecentered{} and \GreedyIt{-l} is between
\num{0.2} and \num{1.5}.
However,
for each value of $\B$,
the average speedup across all instances does not exceed \num{0.9}.
\GreedyIt{} is faster than \kEC{} only on the \DSpfab{} dataset with speedups up to \num{1.3},
but takes more than twice as long as \kEC{} on the other datasets.
Only \BatchInvariant{} achieves significant improvements over \kEC{}, with the average speedup over all \DSpfab{} instances ranging from \num{1.3} to \num{1.7}.
On average across all instances, however, \emph{\kEC{} remains faster than all batch-dynamic algorithms}.

In terms of solution weight, the algorithms differ by no more than \SI{2.5}{\percent} from the solution weight of \kEC{}.
We conclude that among the static and batch-dynamic algorithms, \emph{\kEC{} performs best} and that \emph{\BatchInvariant{}} is the only batch-dynamic algorithm that is \emph{competitive}
on some datasets.
\emph{In the following, we hence use \kEC{} as reference for comparisons.}

\paragraph*{Dynamic Algorithms}
We compare \dynGreedy{} and the randomized \dynGreedy{-r} to the respective variants with both filtering and post-processing, \dynGreedy{-pf} and \dynGreedy{-rpf}.
The latter two
achieve  equal solution weight,
which is in all cases at least \SI{98}{\percent} of the weight achieved by \kEC{}.
\dynGreedy{-pf} is always slower than \dynGreedy{-rpf}, which achieves speedups between \num{0.6} and \num{3.3} relative to \kEC{} across all $\B$ and instances.
The close performance in solution weight of \dynGreedy{-pf} and \dynGreedy{-rpf} suggests
that the \emph{post-processing contributes significantly to the solution weight} and can compensate for low-weight solutions produced by the base algorithm.

The randomized \dynGreedy{-r} cannot match \kEC{} in terms of solution weight.
For each $\B$,
the average solution weight
across all instances
in \DSfacebook{}
is below \SI{89}{\percent} relative to \kEC{}.
Also \dynGreedy{} does not provide an advantage over \kEC{}:
it is either significantly slower or its solutions are more than \SI{10}{\percent} worse.
Among the \dynGreedy{} variants, \emph{\dynGreedy{-rpf} thus offers the best trade-off} between solution weight and running time.

The dynamic version of \kEC{}, \dynkEC{},
only becomes competitive with \kEC{} when combined with filtering and post-processing.
Without either its solutions are about \SI{10}{\percent}
worse than that of \kEC{} on average for each $\B$ and it takes up to three times as long as \kEC{}, except for $\B \leq 4$ on \DSfacebook{}, where the speedup is up to \num{1.7}.
Enabling post-processing (\dynkEC{-p}) improves the solution weight
slightly, but makes it even slower.
The combination with filtering (\dynkEC{-pf}) again compensates the loss in running time:
\dynkEC{-pf} is faster than \dynkEC{} overall, but still slower than \kEC{},
however by only \SI{4}{\percent} on average.
Its solution weight is comparable to \dynGreedy{-rpf} (cf.~\Figure{rw-boxplot-tpu}).

\paragraph*{Hybrid Algorithms}
\kEC{} and \hybridkEC{} perform similar in terms of solution weight,
differing in no more than \SI{2}{\percent} on any instance.
\hybridkEC{} also tends to be at least as fast as \kEC{}:
On average over all $\B$ and all instances, it is equally fast on \DSfacebook{},
\SI{40}{\percent} faster on \DSpfab{} and \SI{8}{\percent} faster on \DShpc{}.
Both algorithms do not benefit from post-processing, which increases the running time but not the solution weight.

Hybridizing \kEC{} with 
\dynGreedy{-rpf} yields the algorithm \hybridGreedy{-rpf}, which
produces solutions that are for each $\B$ on average no more than \SI{1}{\percent}
worse.
It is also faster than \kEC{}, with average speedups up to \SI{40}{\percent}.

\paragraph*{Summary}
\kEC{} is clearly the best static algorithm on the real-world instances,
and also superior to all batch-dynamic algorithms except \BatchInvariant{}.
Among the dynamic algorithms, \dynGreedy{-rpf} provides the best performance overall, with large speedups and solutions close in weight to those of \kEC{}.
\hybridGreedy{-rpf} and \hybridkEC{} are also faster than \kEC{}
and have comparable solution weight.

\subsubsection{\underline{Split Instances}}
\begin{figure}
    \centering
    \includegraphics[width=.75\linewidth]{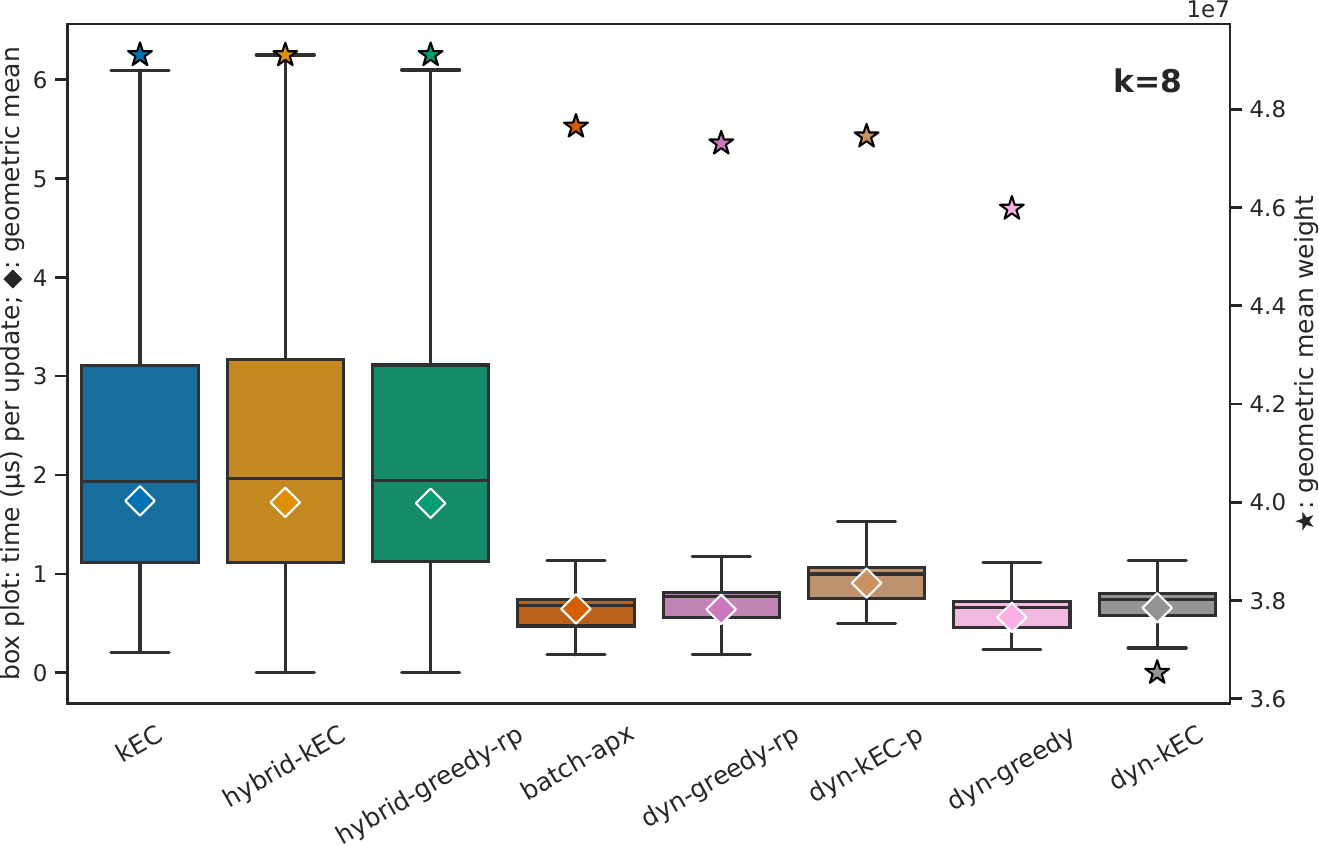}
    \caption{Average per-update times (left axis, boxes and $\blacklozenge$) and weight (right axis, $\star$) for $\B=8$ on all split instances.}\FigLabel{split-boxplot-tpu}
\end{figure}
On the \DSfacebooksplit{} instances, we see a similar, though not identical picture
in comparison to the real-world instances (cf.~\Figure{split-boxplot-tpu}).
One notable difference is that due to the limited maximum edge weight,
\emph{filtering remains essentially without effect}.
As the number of updates per sub-batch decreases
as the number of sub-batches increases or for a smaller weight limit,
the speedup achieved by the dynamic algorithms relative to \kEC{} 
improves in these cases. 

The speedups are more pronounced on average for small $\B$.
The \emph{fastest algorithms} across all values of $\B$ are
\dynGreedy{-rp} and \BatchInvariant{}, both with a mean speedup
across all $\B$ of \num{2.5} and with a maximum speedup over \kEC{}
on an instance of \num{5.1} and \num{6.0}, respectively.
The hybrid algorithms show a similar performance as \kEC{} with a speedup  between \num{0.85} and \num{1.2}.

All algorithms perform similar w.r.t.\ solution weight,
only \dynGreedy{-r} and \dynkEC{} (both without post-processing) fall
significantly behind the others with a weight relative to \kEC{}
in the ranges \numrange{0.7}{0.9} and \numrange{0.5}{0.9}, respectively.

\subsubsection{\underline{Synthetic Instances}}
\begin{figure}
\centering
\includegraphics[width=.75\linewidth]{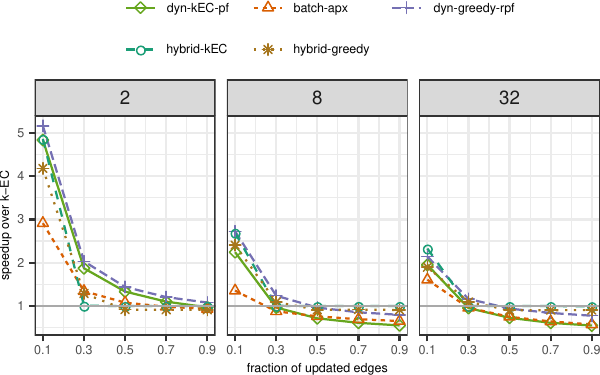}
\caption{Speedup of (batch-)dynamic algorithms over \kEC{} for $\B=2,8,32$ on the synthetic instances,
where fraction of updated edges = \#updates / m.}\FigLabel{speedup-vs-fraction}

\end{figure}
\kEC{} is the fastest static algorithm 
on the synthetic instances.
It is also faster than the batch-dynamic algorithms \BatchGreedy{},
\BatchGreedy{-l}, and \BatchNC{}.
\BatchInvariant{} is faster than \kEC{} if at most \SI{10}{\percent} of the edges are updated, and
for $\B=2$ even if up to half of the edges are updated.
The average speedups of \BatchInvariant{} over \kEC{}
for each $\B$ are between \SI{4}{\percent} and \num{2.9}.

\Figure{speedup-vs-fraction} shows the speedup of (batch-)dynamic and hybrid algorithms over \kEC{}.
The dynamic algorithms are on average faster than \kEC{} if \SI{10}{\percent} of the edges are updated in a batch.
For $\B=2$, the dynamic algorithms \dynGreedy{-rpf} and \dynkEC{-pf} 
achieve a maximum speedup of \num{5.2} and \num{4.0}, respectively,
even if \SI{70}{\percent} of the edges are updated. 
\hybridkEC{} takes on average the same time as \kEC{} unless
only \SI{10}{\percent} of edges are updated.
Here, it is significantly faster
as it uses the dynamic routine to update the solution. Similarly, \hybridGreedy{-rpf}
is faster if $\B=2$ and \SI{10}{\percent} of edges are updated,
but slower if at least half of the edges are updated.

As $\B$ increases, the solution weights generally improve relative to \kEC{}
and, \eg{} for $\B=8$, they are already within \SI{6}{\percent} for all algorithms.
The worst case arises for $\B=2$ and we discuss it next in more detail:
\BatchInvariant{} produces solutions that differ from those of \kEC{} by no more than \SI{3}{\percent} on average.
Both \hybridkEC{} and \hybridGreedy{-rpf} match the solution weight of \kEC{} on instances where they mainly employ the static algorithm.
However, when only \SI{10}{\percent} of the edges are updated the solution weight drops off by \SI{10}{\percent} relative to \kEC{}.
This matches the performance of their dynamic counterparts:
\dynkEC{-pf} and \dynGreedy{-rpf} yield solutions that are at least \SI{10}{\percent} lighter than those of \kEC{} when only \SI{10}{\percent} of edges are updated.
With increasing batch size the solution weight improves, but for \dynkEC{-pf} does not get within less than \SI{8}{\percent} of \kEC{} for $\B=2$.

In conclusion, while the dynamic algorithms are generally faster than \kEC{} on instances with smaller batch size,
the speedup comes at the cost of a somewhat decreased solution weight.
For $\B=2$, this decrease can be quite significant, but it improves for larger $\B$.
\BatchInvariant{} produces high-weight solutions even for $\B=2$,
which are within \SI{3}{\percent} of \kEC{}'s.
Thus, \emph{\BatchInvariant{} is the recommended choice for small $\B$},
whereas \emph{\hybridkEC{} and \hybridGreedy{-rpf} are faster for larger $\B$}
without sacrificing solution weight.

\subsubsection{\underline{Recourse}}
We compare the best algorithms from the previous sections.
As the static algorithms differ in recourse by no more than \SI{2}{\percent},
we use \kEC{} as reference.

\paragraph*{Real-World Instances}
\dynGreedy{-rpf} and \BatchInvariant{} have similar recourse
on \DSfacebook{} and \DShpc{},
which is between \SI{56}{\percent} and \SI{69}{\percent} of \kEC{}'s.
On \DSpfab{}, \BatchInvariant{}
is slightly better than \dynGreedy{-rpf} and
clearly better than \kEC{}.
\dynkEC{-pf} performs similarly well as \dynGreedy{-rpf} on \DShpc{} and \DSpfab{},
and even better on
\DSfacebook{}
with a recourse of at most \SI{39}{\percent} relative to \kEC{}.

The recourse of
the hybrid algorithms
tends to be between \kEC{} and their dynamic subroutine.
\hybridkEC{} performs equal to \kEC{} on \DSfacebook{},
better on \DShpc{}, but worse on \DSpfab{}.
\hybridGreedy{-rpf} has lower recourse than \kEC{} on all instances
and
matches that of \dynGreedy{-rpf} on \DSpfab{}.

\paragraph*{Synthetic Instances}
We expect the recourse of the dynamic and hybrid algorithms to be less than that of \kEC{} for small batch sizes.
The results on the synthetic instances confirm this.
\dynGreedy{-rpf}, \dynkEC{-pf}, and \BatchInvariant{} show similar results in terms of recourse: for $\B=2$ the recourse ranges between \SI{50}{\percent} and \SI{95}{\percent} relative to \kEC{} and between \SI{22}{\percent} and \SI{36}{\percent} for $\B=32$.
As expected, the recourse increases with the batch size.
Surprisingly on first sight,
the recourse relative to \kEC{} decreases as $\B$ increases.
However, for larger $\B$ more edges are colored and, thus, a larger
percentage of edges keep their color during an update.
The hybrid algorithms \hybridkEC{} and \hybridGreedy{-rpf} have the same recourse as \kEC{} if at least half of the edges are updated in a batch.
If only \SI{10}{\percent} of edges are updated,
the recourse is similar to their dynamic counterparts.
This behavior reflects the above discussion of running times and solution weights.

\paragraph*{Split Instances}
We observe a similar reduction in recourse for small batches on the split instances \DSfacebooksplit{}.
\hybridkEC{} and \hybridGreedy{-rp} showed the same performance w.r.t.\ running time and solution weight, and
they also produce the same recourse as \kEC{}, regardless of batch size.
The dynamic algorithms' recourse is on average at least \SI{20}{\percent} lower than that of \kEC{}.
Post-processing has a significant impact on recourse:
for \dynGreedy{-rp} has average recourse up to twice that of \dynGreedy{-r}.
A similar trend can be seen for \dynkEC{-p} compared to \dynkEC{}, but the effect is not as pronounced.
Filtering, on the other hand, has no significant impact on the recourse,
which also matches our observations w.r.t.\ running time and solution weight.
In general, the recourse relative to \kEC{} increases for dynamic algorithms as the batches become larger.
For $\B=2$ \dynGreedy{-rp} produces an average recourse between \SI{51}{\percent} and \SI{73}{\percent} and
\SI{44}{\percent} to \SI{74}{\percent} for $\B=32$.
\dynkEC{-p}, \dynGreedy{-rp} and \BatchInvariant{} have a recourse roughly equal to \dynGreedy{-rp}.

\paragraph*{Summary}
The (batch-)\,dynamic and hybrid algorithms generally
have a smaller recourse
than the static algorithms, which leads to fewer changes in the network configuration.
On instances where the hybrid algorithms mainly use the static \kEC{},
they have no advantage in terms of recourse.

\subsubsection{\underline{Summary of Experiments}}
Our results show that dynamic algorithms are particularly good on instances with smaller batches,
while static algorithms perform well on large batches.
This suggests the following recommendations: (1) When dealing with small batches and small numbers of matchings, i.e., frequent reconfigurations of the network and a small number of switches,
\BatchInvariant{} exhibits the best trade-off between running time and solution weight.
(2) For a larger number of matchings, the dynamic algorithms \dynGreedy{-rpf} and \dynkEC{-pf} provide faster running times and high-weight solutions.
(3) When dealing with large batches, e.g.\ if the intervals between reconfigurations
are longer,
the hybrid algorithms \hybridkEC{} and \hybridGreedy{-rpf} 
provide the best performance if running time, solution weight, and recourse
are considered.
They are also well-suited for other situations and thus
are a very good general-purpose choice.
(4) If the solution weight has the top priority and sacrifices w.r.t.\ running
time and recourse are acceptable, the best choice is \kEC{}.
Generally, the dynamic and hybrid algorithms have a lower recourse than the static algorithms, with \dynkEC{-pf} giving recourse as low as \SI{30}{\percent} relative to \kEC{}.
Note that  reduced recourse  implies less reconfiguration cost in the optical network and, thus, less overhead.

\section{Conclusion}\label{sec:conclusion}
To improve the performance of emerging reconfigurable datacenter networks, we studied efficient
algorithms to maximize the amount of traffic which can be offloaded to the optical
topology in a demand-aware manner.
In particular, we presented several dynamic algorithms which exploit the temporal structure of workloads, by computing the disjoint matchings offered by optical switches in an adaptive manner.

Our work opens several interesting avenues for future research.
In particular, we have so far focused on algorithms running on a centralized controller, which matches the predominant architectural datacenter network model today~\cite{singh2015jupiter,zhang2021gemini,osn21}. Still, it would be interesting to explore  distributed versions of our algorithms: emerging reconfigurable datacenter architectures support a distributed control, e.g., performed directly on the switches.
Furthermore, we have so far focused on \emph{online} algorithms which do not require any knowledge of future traffic demands. This is attractive, as the predictability of such traffic demands is application-specific. Nevertheless, it would be interesting to explore how a dynamic scheduler can benefit from headroom, as it may be available for repetitive applications (e.g., related to learning):
such knowledge may be exploited to pre-compute solutions dynamically to some extent and hence to further improve the algorithms.
Our algorithms may also be adapted to the related problem of scheduling connections in satellite-based communication networks, where colors correspond to time slots~\cite{feige2002approximating}.
The problem differs in that the matchings are unweighted and bounded in cardinality.

\balance 

\bibliographystyle{IEEEtran}
\bibliography{IEEEabrv,manuscript}

\end{document}